# A Leader-Driven Open Collaboration Platform for Exploring New Domains

Michael Weiss
Carleton University
TIM Program
Ottawa, ON K1S 5B6
Canada
michael_weiss@carleton.ca

Ibrahim AbuAlhaol
Carleton University
TIM Program
Ottawa, ON K1S 5B6
Canada
ibrahimee@ieee.org

Mohamed Amin
Carleton University
TIM Program
Ottawa, ON K1S 5B6
Canada
mohamed.amin@carleton.ca

## ABSTRACT

This paper describes the design and initial evaluation of a leader-driven open collaboration platform for exploring new domains. The goal of this platform is to enable the collaboration of subject matter experts across knowledge boundaries. Traditionally, new domains are explored from within a single specialist or a focused group perspective. However, this often introduces bias. Collaboration helps reduce such bias by providing access to a broader range of information sources, increasing the chances for producing new insights in a new domain. However, it also introduces a new problem: variance between the contributions made. Variance makes it difficult to produce a coherent document. In this paper, we report on our observations from developing an initial prototype of the open collaboration platform, and derive propositions about how leader-driven open collaboration helps reduce bias while containing variance.

## 1. INTRODUCTION

Today's big challenges (whether climate change, searching for alternative energy sources, or protecting ourselves from cyber attacks) require us to become good at exploring new domains, so that we can find feasible and practical solutions to those challenges. For example, network operators face threats from a wave of new malware releases, as many as 80,000 per day. Their analysts are looking for practical ways to mitigate this massive threat. As part of this task, they need to explore code reuse attacks, since most malware is created by modifying existing malware or by reusing existing techniques to bypass cyber defences. The literature in this domain is large and emerging quickly. There are many types of code reuse attacks. How can a network operator stay on top of the latest types of attacks and prevention techniques?

Traditionally, new domains are explored from within a single specialist perspective (i.e., an organization, a department in an organization, an academic discipline). This perspective influences what kind of information sources are consulted and how the information is synthesized. In other words, the perspective can introduce a bias. Collaboration helps reduce such bias by providing access to a broader range of information sources, thus increasing the chances for producing new insights and opening a new perspectives on the domain. However, collaboration also introduces a new problem: variance between the contributions made. This makes it difficult to produce a coherent document.

In this paper, we describe the design and initial evaluation of a leader-driven open collaboration platform for exploring new domains. The goal of this platform is to enable the collaboration of subject matter experts across knowledge boundaries. The term "knowledge boundary" is used by [1] to describe the social boundaries that structure knowledge, whether they are disciplines, functions, or organizations. The initial design of the platform was first described in [2]. Here, we offer lessons from a prototype of the platform and derive propositions about the approach.

The article first provides additional background on exploring new domains, the structure of literature reviews, and prior work on collaborative writing. It then presents the design of a leader-driven open collaboration platform for exploring new domains. This section is followed by observations on an initial prototype of the open collaboration platform, and propositions about how the platform helps reduce bias while containing variance. In the final section, we present our conclusions and describe future work.

## 2. BACKGROUND

In this section, we provide additional background on exploring new domains, the literature reviews, and collaborative writing.

### 2.1. Exploring new domains

A standard approach to exploring a domain is to conduct a literature review [3]. However, conducting a literature review in a new domain presents unique challenges. Whereas in an existing domain, researchers can use established classifications of knowledge to guide their search for and interpretation of the literature, this is not the case for a new domain that lacks such classifications. The task of the researcher is to make sense of evidence when it does not fit existing models and classifications and to extend existing knowledge accordingly.

Exploring a new domain can be thought of as looking for anomalies in the evidence that cannot be explained by what is already known, and subsequently building models and classifications that incorporate this evidence [2]. A particular challenge in exploring a new domain is that the very criteria for searching the domain are co-evolving with our understanding of the domain. At the outset of the literature review, there are few established criteria for what the researchers should be looking for. Lin et al. [4] refer to as a "needle in a haystack problem where the appearance of the needle is unknown".

Attenberg et al. [5] observed that organizations make decisions using established predictive models creating a blind spot . We conclude that, in a new domain, researchers should particularly be looking for areas in the existing knowledge that are supposedly firmly established. In [5], Attenberg et al. suggest to involve non-experts with an outsider's perspective to identify knowledge gaps. These correspond to situations where the model is confident but wrong (these are "unknown unknowns"), not where the model is uncertain ("known unknowns").

## 2.2. Structure of a literature review

The goal of a literature review is to synthesize the current knowledge on a given topic based on previously published research. Creating a literature review involves searching through the literature, retrieving sources of information, and synthesizing the findings of those sources into one paper [3]. With [3], we can differentiate three broad categories of literature reviews: narrative literature reviews, qualitative systematic literature reviews, and quantitative systematic literature reviews. Given the fragmented and evolving nature of the literature in a new domain, the type of literature review most suitable for exploring a new domain is a cross between a narrative and qualitative systematic literature review. Green et al. describe a systematic narrative literature review as a "best-evidence synthesis". The elements of a systematic narrative literature review are shown in Table 1.

**Table 1. Elements of a systematic narrative literature review.**

| | |
|---|---|
| **Focus** | The authors should state the purpose or focus of the literature review. |
| **Relevance** | The authors also need to make a case for the relevance of the review. |
| **Glossary** | The literature review should define any unusual terminology. |
| **Sources of information** | The authors of the literature review need to report on the databases searched and the keywords used. |
| **Search terms** | To limit the number of papers that need to reviewed, the authors should turn the main concepts of the explored domain into search terms. |
| **Selection criteria** | The literature review should describe on what grounds papers were included or excluded. Such criteria help avoid bias in paper selection. |
| **Synthesis** | The information obtained from the literature should be organized into common streams. Tables are a good way of categorizing the evidence collected. A goal of the synthesis is to identify agreements, disagreements, and gaps in the literature. |
| **Limitations** | The authors should identify weak points of the review and areas for future work. |
| **Conclusion** | The conclusion should relate back to the focus and summarize the major findings of the literature review and identify its contributions to knowledge. |

## 2.3. Collaborative writing

Several recently proposed systems for collaborative writing use a crowdsourcing approach to complete the writing task [6, 7]. Crowdsourcing is a technique for leveraging a group of collaborators to solve complex problems [8]. In crowdsourcing, there are two types of users: requesters and members of the crowd [8]. Requesters are the people or organizations who define a problem or task, and aggregate the partial solutions produced by the crowd. Crowd members are people who contribute.

The Ensemble system [6] is based on a cognitive model of writing in [9]. This model views the writing process as a series of rhetorical problems. For each writing task, there is a top-level rhetorical problem that includes the constraints given to the writers and the goals the writers create for themselves. This top-level problem can be further decomposed into sub-problems. The rhetorical problems frame the writing task.

In the MicroWriter system, the task of writing is decomposed into three types of subtasks: generation of ideas, labeling ideas to identify groups of related ideas, and writing paragraphs from related ideas [7]. A key insight is that each subtask should be completed with limited awareness of what has been done already and what others are doing. To this end, the context for completing a subtask is embedded within the subtask itself.

# 3. DESIGN OF A LEADER-DRIVEN COLLABORATION PLATFORM

The design of our leader-driven collaboration platform is modeled on previous work on leader-driven collaborative writing in [6] and crowd-based categorization of documents in [10]. In a leader-driven approach to collaborative writing, there are two types of participants: leaders, who constrain and specify the nature of the contributions – the lead author of a literature review sets the scope of the literature review and guides the synthesis process, and contributors, who are recruited to focus on specific writing tasks.

Following the cognitive model in [9], we conceptualize creating a literature review as a series of writing tasks. We decompose the task of writing a literature review into subtasks for each section of the literature review. Standard elements to build a literature review are identified in Table 1. However, a more fine-grained decomposition is usually required, e.g. separate sections for each literature stream need to be added. These sections are defined by the lead authors of the literature review.

For each section, the lead author motivates (or justifies) the need for the section and specifies a prompt or question (such as "define key features of topic X", or "identify examples of X") that helps focus the contributors' work. Contributors, as well as the lead author, provide alternative drafts in reply to the question. Finally, contributors or lead authors can comment on and categorize those drafts. It is up to the leader to choose the best draft for each section, in order to produce a final version of the article. Often, leaders will end up writing their own drafts in response to their questions, but build on the drafts contributed by others.

Figure 1 shows the workflow supported by the collaboration platform. At the collection stage, the leader solicits contributions from contributors in the form of drafts that address specific questions. At the staging stage, the leader selects the drafts to include into the article. At the synthesis stage, the leader composes an article from the selected drafts. This article is revised at the refinement stage. While the figure suggests a strictly linear flow, in actual use this will be an iterative process, where drafts can be staged as soon as drafts for some of the sections have been collected, a tentative article can be composed by synthesizing those drafts, and the creating of the final article involves the leader editing drafts, asking for additional information in the form of new drafts as necessary, and organizing the content.

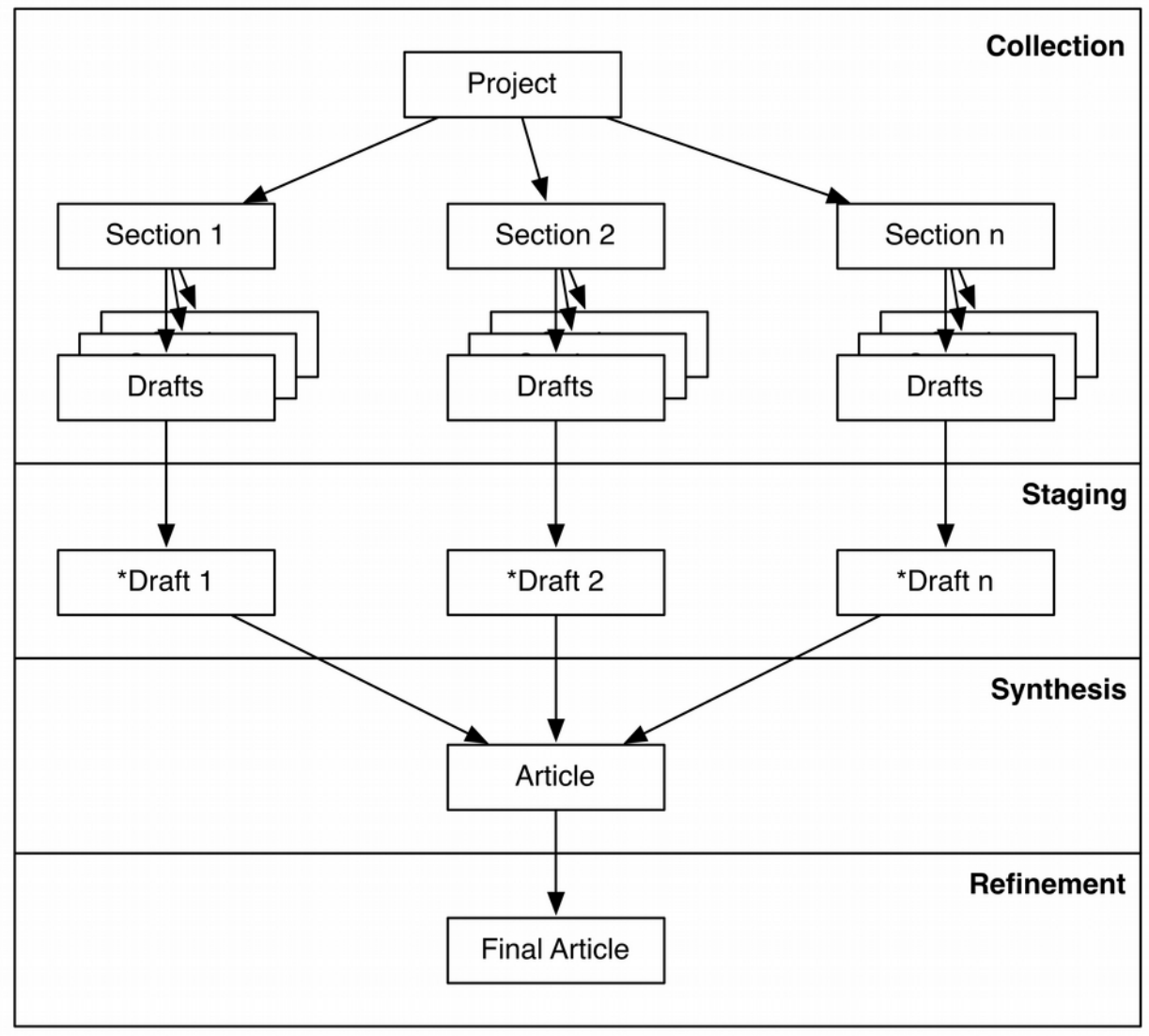

**Figure 1. Workflow implemented by the platform**

# 4. OBSERVATIONS AND PROPOSITIONS

In this section, we report on our observations on an initial prototype of the leader-driven open collaboration platform. The observations are based on an experiment with two groups of researchers conducting a literature review using the collaboration platform. Members of the team were asked for their impressions on using the prototype and how it helped carry out a literature review. Building on these observations we developed propositions about how the leader-driven open collaboration platform helps reduce bias while containing variance.

## 4.1. Reducing bias

Bias can blindside an organization to miss important evidence. Existing knowledge in one area can prevent an organization from accessing knowledge in another area. Prior knowledge has been shown to limit the ability of entrepreneurs to recognize business opportunities in other areas [11]. An organization's established predictive model can cause it to ignore evidence that contradicts the predictive model [5]. As noted by Green et al. [3], literature reviews may be biased by the researcher's perspective, both in terms of the literature included and the synthesis (conclusions drawn from the literature). One way of reducing the bias is to conduct a more rigorous, systematic literature review. Bias can also be reduced by increasing the diversity of perspectives.

Open collaboration allows collaborators to share information sources and to build on each other's findings. The diversity (in terms of area of expertise, culture, etc.) of the collaborators helps reduce bias that may exist when the research is conducted by a single person. The literature on team diversity predicts that increasing knowledge diversity in a team positively affects the range of information accessible to it [12]. In the experiment, contributors read sources contributed by others and challenged their interpretations by commenting and replying to comments.

***H1a:*** *Open collaboration helps collaborators find and interpret information sources in the new domain.*

A significant task during the exploration of a new domain is to group the knowledge into research streams, and to identify research gaps. In [10], crowd members iteratively categorize text fragments. When a crowd member is asked to categorize a fragment, they see how other fragments have been categorized. They can then decide to put the new fragment into an existing category or create a new one. In our experiment, one leader asked for examples of code reuse attacks. As contributors collected and categorized the examples, they produced a taxonomy of code reuse attacks, including one novel type of code reuse attack.

***H1b:*** *Open collaboration helps collaborators organize the knowledge in a new domain.*

Researchers provide support for the relevance of a research problem as well as the validity of its solution in the form of evidence [3]. According to [13], evidence has two attributes: type (e.g., prior literature, presentations, examples), and validity (i.e., following a research methodology, authorization by experts, and support by a large community). The authors in [14] describe a crowdsourced process for evaluating literature which involves decomposing a research question into subtasks that can be distributed to the crowd. Participants in the experiment reported that the platform helped them explore different types of evidence that included papers, related news websites, videos, and online lectures. It also helped them identify the key researchers in the problem domain, lending credibility to the evidence.

***H1c:*** *Open collaboration helps collaborators find and organize evidence, in particular non-traditional evidence.*

Creating a taxonomy is an important starting point for conducting research in a new domain. It provides context and helps direct the attention of researchers to relevant problems. In the experiment, the taxonomy helped participants identify different definitions of the "anticipation" concept. It also helped them understand the relationship between anticipation and prediction and articulate the differences between those concepts. In the collaboration platform, comments were used to categorize concepts and link them.

***H1d:*** *Open collaboration helps collaborators create a taxonomy of concepts and their relationships in a new domain.*

## 4.2. Containing variance

While helping contain bias, collaboration introduces a new problem: variance between the contributions due to stylistic differences and inconsistencies in content. This variance makes it difficult to produce a coherent document. Misaligned contributions from different authors can impair the consistency of a document and make it harder to read and understand [15]. In their work, Silva & Henderson consider a document to be coherent, if there is a "smooth and natural progression of ideas between them" (i.e., when it conveys a consistent narrative).

In leader-driven open collaboration, there are two types of participants: leaders and contributors. Leaders are responsible for the overall vision and flow of the document, while contributors provide input in their specific areas of expertise [6]. Leaders and contributors thus have complementary motivations. In our experiment, leaders were generally quite opinionated about the structure and direction of the articles.

***H2a****: Collaborators adapt to the asymmetric collaboration structure.*

Coherence is an important quality of a document that engages readers. It comprises consistency in style and a logical flow of argument [15]. When the argument flows logically, reading the document raises questions in the reader's mind which are subsequently answered, increasing the retention of the message to be conveyed by the document. In the experiment, both leaders reported that having a list of section templates to choose from helped them maintain a clear structure. We also observed that leaders synthesized contributed drafts into final drafts. However, to firmly conclude that a leader-driven approach increases coherence, we need to operationalize the coherence construct and evaluate a larger sample of articles.

***H2b:*** *Leader-driven open collaboration allows collaborators to create more coherent documents than leader-less collaboration.*

We expect that leaders seek and incorporate feedback (in the form of drafts and comments) from contributors. In the Ensemble system in [6], leaders reported that they found the perspectives of others beneficial. In our experiment, leaders often gave specific directives to contributors on what kind of feedback they sought, for example, "identify the reasons why anticipation is critical in cybersecurity referring to examples in the popular press".

***H2c:*** *Feedback from contributors can improve the content and quality of the article created, but it can also lead to dispersed ideas and disagreement.*

User engagement and retention is important for the adoption of a collaboration platform [16]. Contributors need to perceive that their contributions have impact. Likewise, leaders need to perceive that they succeeded in orchestrating the production of quality and content-rich articles. In the experiment, contributors felt that their contributions were highly valued. Leaders and contributors effectively collaborated on reducing variance.

***H2d:*** *Contributors perceive that their expertise is valued and leaders are able to harness the contributions from contributors.*

# 5. CONCLUSIONS

In this paper, we described a leader-driven approach to open collaboration for the exploration of new domains. We presented observations on an initial prototype of the platform and derived a set of propositions. Our initial observations lend tentative support for our hypothesis that a leader-driven approach can help reduce single perspective bias while containing the variance between contributions. In the next step of our research, we will conduct a formal experiment to test the propositions.

# 6. ACKNOWLEDGMENTS

We would like to recognize the contributions of the other members of the implementation team (Chris Budiman, Raed Iskander, Ali Abu Alhawa), and the researchers involved in conducting the experiment (Mahmoud Gad, Ahmed Shah) by including them as co-authors of this paper.